\documentclass[sigconf,natbib=true]{acmart}
\usepackage{setspace}
\usepackage{amsmath}               
  {
      \theoremstyle{plain}

  }
\usepackage{enumitem}
\AtBeginDocument{%
  \providecommand\BibTeX{{%
    \normalfont B\kern-0.5em{\scshape i\kern-0.25em b}\kern-0.8em\TeX}}}

\copyrightyear{2024}
\acmYear{2024}
\setcopyright{rightsretained}
\acmConference[CIKM '24]{Proceedings of the 33rd ACM International Conference on Information and Knowledge Management}{October 21--25, 2024}{Boise, ID, USA}
\acmBooktitle{Proceedings of the 33rd ACM International Conference on Information and Knowledge Management (CIKM '24), October 21--25, 2024, Boise, ID, USA}
\acmDOI{10.1145/3627673.3680036}
\acmISBN{979-8-4007-0436-9/24/10}

\author{Yaochen Zhu}
\affiliation{%
\institution{University of Virginia}
\city{Charlottesville}
\state{VA}
\country{USA}
}
\email{uqp4qh@virginia.edu}

\author{Liang Wu}
\affiliation{%
\institution{LinkedIn Inc.}
\city{Sunnyvale}
\state{CA}
\country{USA}
}
\email{liawu@linkedin.com}

\author{Binchi Zhang}
\affiliation{%
\institution{University of Virginia}
\city{Charlottesville}
\state{VA}
\country{USA}
}
\email{epb6gw@virginia.edu}

\author{Song Wang}
\affiliation{%
\institution{University of Virginia}
\city{Charlottesville}
\state{VA}
\country{USA}
}
\email{sw3wv@virginia.edu}

\author{Qi Guo}
\affiliation{%
\institution{LinkedIn Inc.}
\city{Sunnyvale}
\state{CA}
\country{USA}
}
\email{qguo@linkedin.com}

\author{Liangjie Hong}
\affiliation{%
\institution{LinkedIn Inc.}
\city{Sunnyvale}
\state{CA}
\country{USA}
}
\email{liahong@linkedin.com}

\author{Luke Simon}
\affiliation{%
\institution{LinkedIn Inc.}
\city{Sunnyvale}
\state{CA}
\country{USA}
}
\email{lsimon@linkedin.com}

\author{Jundong Li}
\affiliation{%
\institution{University of Virginia}
\city{Charlottesville}
\state{VA}
\country{USA}
}
\email{jundong@virginia.edu}

\usepackage{tikz}
\usepackage{subcaption}
\usepackage{pgfplots}
\usetikzlibrary{arrows,positioning,automata,calc,shapes}
\pgfplotsset{compat=newest, scaled z ticks=false} 
\pgfplotsset{plot coordinates/math parser=false}
\newlength\figureheight 
\newlength\figurewidth
\usepackage{tikz}
\usepackage{caption}
\usepackage{filecontents}
\usepackage{url}
\usepackage{amsmath}
\usepackage{graphicx}
\usepackage{url}
\usepackage{hyperref}

\usepackage{amsfonts,amssymb}
\usepackage{color, soul}
\usepackage{mathrsfs}
\usepackage{cleveref}
\usepackage{multirow}
\usepackage{wrapfig}
\usepackage{amsthm}
\usepackage{algorithm}
\usepackage{algorithmic}
\usepackage{bm}
\usepackage{mdframed}

\begin{document}

\fancyhead{}

\title{Understanding and Modeling Job Marketplace with\\ Pretrained Language Models}

\begin{abstract}

Job marketplace is a heterogeneous graph composed of interactions among members (job-seekers), companies, and jobs. Understanding and modeling job marketplace can benefit both job seekers and employers, ultimately contributing to the greater good of the society. However, existing graph neural network (GNN)-based methods have shallow understandings of the associated textual features and heterogeneous relations. To address the above challenges, we propose PLM4Job, a job marketplace foundation model that tightly couples pretrained language models (PLM) with job market graph, aiming to fully utilize the pretrained knowledge and reasoning ability to model member/job textual features as well as various member-job relations simultaneously. In the pretraining phase, we propose a heterogeneous ego-graph-based prompting strategy to model and aggregate member/job textual features based on the topological structure around the target member/job node, where entity type embeddings and graph positional embeddings are introduced accordingly to model different entities and their heterogeneous relations. Meanwhile, a proximity-aware attention alignment strategy is designed to dynamically adjust the attention of the PLM on ego-graph node tokens in the prompt, such that the attention can be better aligned with job marketplace semantics. Extensive experiments at LinkedIn demonstrate the effectiveness of PLM4Job.

\end{abstract}

\ccsdesc[500]{Information systems~Data mining\vspace{-0.25ex}}

\keywords{Large Language Model; Graph Mining; Job Marketplace}

\maketitle

\section{Introduction}
Job marketplace is a pivotal component of our society \cite{borisyuk2017lijar,du2024enhancing}. A job marketplace can be viewed as a heterogeneous graph of members (job seekers), jobs, and companies, where companies release job postings that members can apply for, and members establish social connections by following one another. In general, job postings contain job descriptions and requirements to recruit suitable talents, whereas members are typically associated with abundant self-provided textual features such as biographies, skills, experiences, etc., to increase the likelihood of securing good employment.

Various interesting tasks can be conducted on job marketplace to contribute to the welfare of its stakeholders. From the entity's perspective, since some members do not provide certain important attributes (e.g., skills) in their profile, member attribute prediction is essential to more precisely match them to potential job opportunities \cite{jiechieu2021skills,zhu2023path}. Additionally, with the recent COVID epidemic, there is growing interest in predicting members' work mode preference (e.g., onsite, online, or hybrid), such that job postings can be pre-filtered to save limited recommendation budgets \cite{zhu2022remote}. For the relational level task, it is beneficial to suggest members with other members to follow \cite{geyik2019fairness} or suitable jobs to apply for \cite{zhu2022mutually,zhu2024collaborative,zhu2022variational}.

Graph neural networks (GNN) \cite{wu2020comprehensive} can be used to model the job marketplace to tackle the aforementioned tasks \cite{shalaby2017help,de2021job,guo2024consistency}. For example, \citet{zhu2022remote} propose to model member-job interactions as a bipartite graph to predict members' work mode preference given their interacted jobs. In addition, \citet{wang2023improved} propose a heterogeneous GNN for job recommendations. Nevertheless, GNN-based approaches lack prior knowledge of the diverse member-member relations (e.g., follow, co-work) or member-job relations (e.g., follow, apply, view) in the job marketplace, resulting in a limited understanding of the relationships among different entities. In addition, since most GNN-based methods adopt bag-of-word relations of the rich textual data associated with members and jobs, their understanding of textual information is unavoidably shallow. 

Recently, more efforts have been devoted to using pretrained language models (PLMs) to tackle text-attributed graphs (TAG) \cite{jin2023large,wang2023knowledge,wu2024exploring,ren2024survey}, where their encoded knowledge and reasoning ability can be fully utilized to understand the node textual features and their relations \cite{wu2024usable,liu2024knowledge}. The key challenge is to introduce graph structures to PLMs. Generally, there are two strategies to address the issue. One main strategy is to integrate auxiliary GNNs with PLMs, which either views the PLMs as node feature extractors \cite{zhu2021textgnn}, or uses GNN embeddings (projected into the PLM token embedding space) to represent the nodes in the PLM \cite{tian2023graph,tang2023graphgpt}. However, these strategies unavoidably inherit the drawbacks of the auxiliary GNN and introduce extra computational overhead. Another strategy is to use natural language to describe the proximity relationship between nodes in the graph, e.g., using textual descriptions such as "\texttt{node\_1 and node\_2 are within \textbf{one-hop}}" to denote the connected relation between "\texttt{node\_1}" and "\texttt{node\_2}" \cite{ye2023natural}. However, since there is no evidence that such textual descriptions can properly guide the PLM to attend to the nodes based on the proximity relation in the graph, the graph structure is still loosely coupled with the PLM.

To address the above challenges, we propose a graph-oriented PLM, i.e., PLM4Job, to tightly couple the pretrained knowledge with the heterogeneous structure of the job market graph, which could serve as the foundation model for various downstream tasks on the job marketplace. Specifically, we first introduce member/job tokens to faithfully represent nodes in the job marketplace graph. Then, in the pretraining phase, we propose a novel heterogeneous ego-graph-based prompting strategy to model and aggregate member/job textual features based on the topological structure around the target member/job, where entity embeddings and graph positional embeddings are introduced accordingly to facilitate the PLM to understand various entities and their respective relationships in the job marketplace. In addition, a proximity-aware attention alignment strategy is introduced to dynamically adjust the attention of the backbone PLM on the ego-graph node tokens in the prompt, such that the attention of the PLM can be better aligned with job marketplace semantics. Finally, for node-level tasks, we introduce label tokens for efficient, hallucination-free predictions. 

\section{Methodology}
In this section, we introduce the problem setting of treating the job marketplace as a heterogeneous text-attributed graph (TAG) and the proposed PLM4Job as a foundation model to tackle various entity-level and link-level downstream tasks.

\begin{figure}[t]
\centering 
\includegraphics[width=0.42\textwidth]{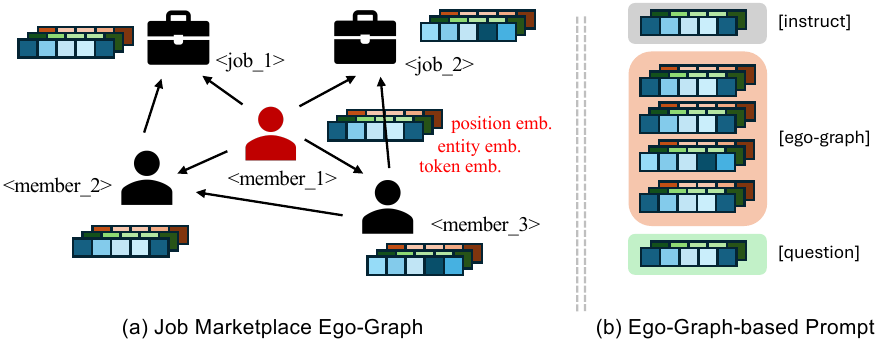}
\vspace{-2mm}
\caption{The job marketplace heterogeneous ego-graph and the corresponding ego-graph-based prompt.}
\label{fig:overview}
\vspace{-3mm}
\end{figure}

\subsection{Problem Formulation}

Suppose we have a job marketplace with a set of members $\mathcal{U} = \{1,2,\ldots, N_{U}\}$ and jobs $\mathcal{I} = \{N_{U}+1,N_{U}+2,\ldots, N_{U}+N_{I}\}$. Generally, each member or job is associated with rich textual features, such as biography, skills, résumé from the member side, and job descriptions (JD) from the job side. In addition, various relationships can be formed among members and jobs. For example, members can apply, click, and view job postings, where the observed relations can be recorded as $\mathcal{R}_{UI} \subseteq \{u \rightarrow i \ | u \in \mathcal{U}, i \in \mathcal{I} \}$. In addition, members can follow each other to form a professional social network, i.e., $\mathcal{R}_{UU} \subseteq \{u \rightarrow u^{\prime} | u, u^{\prime} \in \mathcal{U}\}$. Finally, member $i$ is also associated with certain attributes of interests, which we denoted as $y_{i}$. Here, if we use $\mathcal{N} = \mathcal{U} \cup \mathcal{I}$ and $\mathcal{E} = \mathcal{R}_{UI} \cup \mathcal{R}_{UU}$ to denote the node and edge sets, and use $\mathcal{A} = \{U, I\}$ and $\mathcal{R} = \{R_{UI}, R_{UU}\}$ to denote the entity and relation sets, we can find that the job marketplace can be viewed as a text-attributed heterogeneous graphs $G = (\mathcal{N}, \mathcal{E}, \mathcal{A}, \mathcal{R})$. 

Given that both node attributes and links can be missing from job market heterogeneous graph $G$, the objective of this paper is to understand and model $G$ with graph-oriented pretrained language models (PLM), fully utilizing both graph structure and textual information to make strategic decisions that benefit all the stakeholders.

\subsection{Ego-Graph-based Prompting}
\label{sec:prompt}
In this sub-section, we introduce an ego-graph-based prompting strategy to tightly couple the PLM with the heterogeneous ego-graph $G_{k}$ of target node $k$. An overview is illustrated in Fig. \ref{fig:overview}.

\subsubsection{\textbf{Node Token for Heterogeneous Ego-Graph}}

For PLMs to understand $G_{k}$, we need first tokenize it into a sequence. Since the vocabulary of PLMs may not be able to faithfully represent graph nodes \cite{ye2023natural}, we extend its vocabulary with node tokens and introduce learnable feature embeddings $\mathbf{Z} \in \mathbb{R}^{N \times K}$ to encode node features. Here, we use special tokens with bracket, e.g., "\texttt{<member\_$u$>}", "\texttt{<job\_$i$>}" to denote the newly introduced node tokens for member $u$ and job $i$, respectively. In addition, to faithfully represent the entity and structural information in $G_{k}$, we introduce entity embeddings $\mathbf{E} \in \mathbb{R}^{|\mathcal{A}| \times K}$ and ego-graph positional embeddings $\mathbf{P} \in \mathbb{R}^{(D+1) \times K}$, where $D$ is the maximum depth of $G_{k}$. The final token embedding for node $i \in G_{k}$ can be formulated as follows:
\begin{equation}
\label{eq:token_embds}
\mathbf{h}_{i}^{(0)} = \mathbf{z}_{i} + \mathbf{e}_{A_{i}} + \mathbf{p}_{dist(i, k)},
\end{equation}
where $dist(i, k)$ denote the shortest distance between node $i$ and center node $k$. By converting each node in the heterogeneous ego-graph $G_{k}$ into a token sequence with Eq. (\ref{eq:token_embds}), node features and heterogeneous topological relationships can be well captured. 

\subsubsection{\textbf{Feature Modeling}}

With the ego-graph node tokens and embeddings, we introduce the ego-graph-based prompting strategy to effectively learn the member/job token embeddings via language modeling (LM). We first discuss the feature learning step, which aims to encode the member/job textual features (e.g., member biographies and job descriptions) into the token embeddings. 

\noindent W.L.O.G., if member $k$ is the center node and biography is the textual feature, we first establish the prompt-completion pair $\left(\mathbf{x}_{k}^{f}, \mathbf{y}_{k}^{f}\right)$ as:

\vspace{2mm}
\begin{mdframed}[backgroundcolor=black!10] 
\textbf{Feature Modeling} \\
\vspace{-2.8mm}

\noindent \textbf{\small (a) Ego-graph-based prompt $\mathbf{x}^{f}_{k}$:}\\
\noindent \colorbox{green!50}{\texttt{\footnotesize Given an ego-network in a job marketplace: }} \footnotesize($\mathbf{x}^{ins}$) \ \texttt{\footnotesize $\underset{\footnotesize \mathrm{\textbf{center}}}{\colorbox{red!35}{<member\_$k$>}}$ \\ $\underset{\footnotesize \mathrm{\textit{one-hop}}}{\colorbox{blue!35}{<member\_$i$> <member\_$j$>}}$ $\underset{\footnotesize \mathrm{\textit{one-hop}}}{\colorbox{yellow!50}{<job\_$l$>}}$ $\underset{\footnotesize \mathrm{two-hop}}{\colorbox{yellow!50}{<job\_$m$>}}$ $\underset{\footnotesize \mathrm{two-hop}}
{\colorbox{blue!35}{<member\_$o$>}}$ \footnotesize($\mathbf{x}^{g}_{k}$),\\ \colorbox{green!50}{the biography of the center member} $\underset{\footnotesize \mathrm{\textbf{center}}}{\colorbox{red!35}{<member\_$i$>}}$ \colorbox{green!50}{is}:} $(\mathbf{x}_{k}^{f, q})$\\
\vspace{1mm}
\noindent \textbf{\small (b) Completion $\mathbf{y}^{f}_{k} $:}\\
\noindent \colorbox{green!50}{\texttt{\footnotesize A hard-working applied research scientist at LinkedIn.}}
\vspace{0.5mm}
\end{mdframed}
\vspace{2mm}
Here, we use color {\color{blue} \textbf{blue}} to denote member tokens, {\color{yellow}\textbf{yellow}} to denote job tokens, {\color{green} \textbf{green}} to denote textual tokens, respectively. Ego-graph positional embeddings are denoted with sub-annotation. Specifically, the ego-graph-based prompt for member feature, i.e., $\mathbf{x}^{f}_{k}$, is composed of three parts: \textbf{\textit{(i)}} instruction part $\mathbf{x}^{ins}$, which provides context regarding the job marketplace; \textbf{\textit{(ii)}} ego-graph part $\mathbf{x}^{g}_{k}$, which includes the center node $k$ and a randomly sub-sampled $D$-hop neighborhood as the job marketplace context; \textbf{\textit{(iii)}} question part $\mathbf{x}_{k}^{f, q}$, which naturally leads to the completion $\mathbf{y}^{f}_{k}$.

We use causal language modeling \cite{radford2019language} to learn the ego-graph token embeddings with prompt-completion pairs $\left(\mathbf{x}^{f}_{k}, \mathbf{y}^{f}_{k}\right)$. Specifically, we denote the backbone PLM with extended ego-graph tokens as $P_{\Theta}(x_{t} | \mathbf{x}_{<t})$, which generates the next token $x_{t}$ based on the context token sequence $\mathbf{x}_{<t}$. The parameters $\Theta = \{\hat{\boldsymbol{\theta}}, \mathbf{Z}, \mathbf{E}, \mathbf{P} \}$ are composed of the pretrained PLM weights $\hat{\boldsymbol{\theta}}$ (which is frozen) and the newly introduced embeddings $\Theta_{hot} = \{\mathbf{Z}, \mathbf{E}, \mathbf{P} \}$. The loss of the feature modeling step for PLM4Job can be formulated as follows:
\begin{equation}
\label{eq:feature_lm}
\mathcal{L}^{f}_{k} =  \sum_{t=1} \log \left(P_{\Theta}\left(y^{f}_{k,t} \mid \mathbf{x}_{k}^{f}, \mathbf{y}^{f}_{k,<t}\right)\right).
\end{equation}
Since the completion $\mathbf{y}^{f}_{k}$ contains only textual tokens, when optimizing the ego-graph token embeddings $\Theta_{hot}$ according to Eq. (\ref{eq:feature_lm}), we only calculate the softmax over all the textual tokens, where the stability of language modeling can be substantially enhanced \cite{zhu2024collaborative}.

\subsubsection{\textbf{Metapath-based Structural Modeling}}
\label{sec:struct}

After encoding the node textual features into the corresponding member and job token embeddings according to Eq. (\ref{eq:feature_lm}), we further aggregate the information based on the local job marketplace topology. Here, we define the metapath in a heterogeneous graph $G$ as follows: 
\begin{definition} 
\textbf{Metapath \cite{sun2011pathsim}.} A metapath $\phi$ is defined as a path in the form of $
A_1 \xrightarrow{R_1} A_2 \xrightarrow{R_2} \cdots \xrightarrow{R_l} A_{l+1}$, where $A_{i} \in \mathcal{A}$ and $R_{j} \in \mathcal{R}$ denote the entity and relation, respectively. The metapath can be abbreviated as $A_1 A_2 \cdots A_{l+1}$ with composite relationship $R_1 \circ R_2 \circ \cdots R_l$, where $\circ$ denotes composition operation on relations.
\end{definition}
In the metapath-based structural modeling step, given a predefined set of candidate metapaths $\Phi = \{\phi_{1}, \phi_{2}, \cdots \phi_{M}\}$, for a center node $k$, we aim to transform each compatible metapath $\phi \in \Phi$ (compatible means $A^{\phi}_{1} = \mathrm{type}(k)$) into an ego-graph-based prompt $\mathbf{x}^{\phi}_{k}$, with completion $\mathbf{y}^{\phi}_{k}$ constructed from a randomly shuffled sequence of the end entity $A^{\phi}_{ l+1}$. Then $\mathbf{y}^{\phi}_{k}$ is predicted based on $\mathbf{x}^{\phi}_{k}$ via language modeling. Through this strategy, information in the job market graph can be aggregated along the selected metapaths. The simplest $\phi$ is one-hop metapath, i.e., $\phi \in \Phi_{1} = \{UU, UI, IU\}$. Here, we take the metapath $\phi = UU$ as an example, where the prompt, completion pair $\left(\mathbf{x}^{UU}_{k}, \mathbf{y}^{UU}_{k}\right)$ can be formulated as follows:

\vspace{2mm}
\begin{mdframed}[backgroundcolor=black!10] 
\textbf{First-order Structural Modeling} \\
\vspace{-2.5mm}

\noindent \textbf{\small (a) Ego-graph-based prompt $\mathbf{x}^{UU}_{k}$:}\\
\noindent \colorbox{green!50}{\texttt{\footnotesize Given an ego-network in a job marketplace: }} \footnotesize($\mathbf{x}^{ins}$) \ \texttt{\footnotesize {\colorbox{red!35}{<member\_$k$>}} \\ 
{\colorbox{blue!35}{<member\_$i$> <member\_$j$>}} {\colorbox{yellow!50}{<job\_$l$> <job\_$m$>}}
{\colorbox{blue!35}{<member\_$o$>}} \footnotesize($\mathbf{x}^{g}_{k}$),\\ \colorbox{green!50}{the center member} \colorbox{red!35}{<member\_$k$>} \colorbox{green!50}{follows these members:}} $(\mathbf{x}^{q, UU}_{k})$\\

\noindent \textbf{\small (b) Completion $\mathbf{y}^{UU}_{k}$:}\\
\noindent {\colorbox{blue!35}{\texttt{<member\_$r$> <member\_$s$> <member\_$t$>}}}
\end{mdframed}
\vspace{2mm}

\noindent Here, we note that the question part $\mathbf{x}^{q, UU}_{k}$ of the ego-graph-based prompt $\mathbf{x}^{UU}_{k}$ specify the last relation $R^{\phi}_1$ (i.e., ``follows'') in the metapath $\phi$, such that the encoded knowledge of the PLM can be fully utilized to facilitate the understanding of the relation and predict $y^{UU}_{k}$. In addition, only nodes not selected in the ego-graph $G_{k}$ (i.e., not in $\mathbf{x}^{g}_{k}$) will be sampled in the completion $\mathbf{y}^{UU}_{k}$, which avoids the short cut of direct repeating nodes in the prompt.

Higher-order metapaths are complicated but are also necessary as they provide shortcuts for message passing among member and job nodes. Here, we use two-hop metapath as an example, where $\Phi_{2} = \{UUU, UIU, IUU, IUI\}$. Previous work such as \cite{ye2023natural} use triples $(A_{1}, A_{2}, A_{3})$ to represent two-hop neighbors as token sequences, but this creates lengthy and redundant prompt due to repetition of intermediate nodes. In this paper, we propose a faster approximation strategy to represent high-order metapaths. Specifically, for center node $k$, we first establish a triple $T^{\phi}_{k} = \left(k, \mathcal{A}^{\phi}_{k,1}, \mathcal{A}^{\phi}_{k,2}\right)$, where $\mathcal{A}^{\phi}_{k,1}$ is the set of randomly sampled intermediate nodes starts from $k$, and $\mathcal{A}^{\phi}_{k,2}$ is a set of end nodes sampled from the union list of the end nodes connected with the intermediate nodes in $\mathcal{A}^{\phi}_{k,1}$ \textit{with repetition} (such that important end nodes can be selected with higher probabilities). Here, we take two-hop metapath $\phi = UIU$ as an example. Based on the triple $T^{UIU}_{k}$, the prompt, completion pair $\left(\mathbf{x}^{UIU}_{k}, \mathbf{y}^{UIU}_{k} \right)$ for $\phi$ can be formulated as follows:
\vspace{2mm}
\begin{mdframed}[backgroundcolor=black!10] 
\textbf{Higher-order Structural Modeling} \\
\vspace{-2.5mm}

\noindent \textbf{\small (a) Ego-graph-based prompt $\mathbf{x}^{UIU}_{k}$:}\\
\colorbox{green!50}{\texttt{\footnotesize Given an ego-network in a job marketplace: }} \footnotesize($\mathbf{x}^{ins}$) \ \texttt{\footnotesize {\colorbox{red!35}{<member\_$k$>}} \\ 
{\colorbox{blue!35}{<member\_$i$> <member\_$j$>}} {\colorbox{yellow!50}{<job\_$l$> <job\_$m$>}}
{\colorbox{blue!35}{<member\_$o$>}} \footnotesize($\mathbf{x}^{g}_{k}$),\\ \colorbox{green!50}{the center member} \colorbox{red!35}{<member\_$k$>} \colorbox{green!50}{is interested in these jobs:} {\colorbox{yellow!50}{<job\_$u$> <job\_$v$> <job\_$w$>}}}$({\color{red}\mathbf{x}^{q, UI}_{k}})$\colorbox{green!50}{\texttt{the following users are also}} \\
\colorbox{green!50}{\texttt{interested in \underline{some of} these jobs:}} $(\mathbf{x}^{q, UIU}_{k})$\\

\noindent \textbf{\small (b) Completion $\mathbf{y}^{UIU}_{k}$:}\\
\noindent {\colorbox{blue!35}{\texttt{<member\_$x$> <member\_$y$> <member\_$z$>}}}
\vspace{0.5mm}
\end{mdframed}
\vspace{3mm}
From the above example, we can find that the ego-graph-based prompt for the metapath $UIU$, i.e., $\mathbf{x}^{UIU}_{k}$, is composed of an extra component $\color{red} \mathbf{x}^{q, UI}_{k}$ that describes the intermediate relationship $UI$ and the sampled final-step entities in $\mathcal{A}^{UIU}_{k,1}$, whereas the final relationship $IU$ is described in the question part $\mathbf{x}^{q, UIU}_{k}$ that begs for completion with $\mathbf{y}^{UIU}_{k}$. Similar prompts can be established based on higher-order metapaths. The language modeling loss of structural modeling for metapath $\phi$ can be formulated as follows:
\begin{equation}
\label{eq:structure_lm}
\mathcal{L}^{\phi}_{k} =  \sum_{t=1}\log \left(P_{\Theta}\left(y^{\phi}_{k,t} \mid \mathbf{x}_{k}^{\phi}, \mathbf{y}^{\phi}_{k,<t}\right)\right).
\end{equation}
Since the completion $\mathbf{y}^{\phi}_{k}$ is composed of either homogeneous member tokens or job tokens, we only calculate the softmax over the member/job token space to stabilize the language modeling process. For PLM with symmetric structure, i.e., the weights of the prediction head are tied with token embeddings (e.g., GPT-2 \cite{radford2018improving}), we also tie the weights of the prediction head with the corresponding member/job embeddings, whereas for other non-symmetric PLMs (e.g., LLaMA \cite{touvron2023llama}), another set of randomly initialized embeddings needs to be introduced as the weights for the prediction head.

\subsection{\textbf{Proximity-Aware Attention Alignment}}
\label{sec:proxi}
Another issue that hinders good modeling of job marketplace with PLMs is the \textbf{misalignment} of attention of the PLM with job market graph topology: When optimizing $\Theta$ according to Eqs. (\ref{eq:feature_lm}),  (\ref{eq:structure_lm}), the PLM needs to attend to the prompt $\mathbf{x}^{\{f,\phi\}}_{k}$ and the already-generated completions $\mathbf{y}^{\{f, \phi\}}_{k,<t}$. However, the attention of the backbone PLM may not be well aligned with the member/job ego-graph $G_{k}$, as it may pay more attention to the recent tokens as for natural language, rather than to the important member/job nodes in the ego-graph $G_{k}$. To address this issue, we propose a proximity-aware attention alignment strategy to dynamically adjust the attention weights calculated by the PLM with proximity relations in the heterogeneous ego-graph $G_{k}$ for both feature and structural modeling. 

Here, the key insight is to view the tokens in the completion for \textit{feature modeling}, i.e., $\mathbf{y}^{f}_{k}$, as associated with the center node $k$ (whereas each token in the completion for structural modeling, i.e., $\mathbf{y}^{\phi}_{k}$, is associated with the node itself), and adjust the weights when attending to \textit{member/job nodes} in the prompt based on their heterogeneous proximity in $G_{k}$. Specifically, the (un-normalized) attention when generating after the $t$-th \textit{token} (assumed to be associated with the $j$-th \textit{node}) on the $t'$-th token in the prompt (assumed to be the $j^{\prime}$-th \textit{node}) can be adjusted as follows:
\begin{equation}
\label{eq:attn_adj}
\alpha_{t t^{\prime}}=\frac{\left(\mathbf{h}_t \mathbf{W}_Q\right)\left(\mathbf{h}_{t^{\prime}} \mathbf{W}_K\right)^T}{\sqrt{d}}+\boldsymbol{\psi}_{j j^{\prime}}^T \mathbf{b},
\end{equation}
where $\mathbf{h}_{\{t, t^{\prime}\}} \in \mathbb{R}^{d}$ is the latent representation of the $\{t, t^{\prime}\}$-th token, $\mathbf{W}_Q$, $\mathbf{W}_K$ are the pretrained query and key matrices of the backbone PLM, respectively, $\boldsymbol{\psi}_{j j^{\prime}} \in \mathbb{R}^{M+1}$ encodes the heterogeneous proximity between the node $j$ and the attended node $j^{\prime}$, and $\mathbf{b} \in \mathbb{R}^{M+1}$ is the newly introduced learnable parameters. Specifically, given an ordered set of metapaths $\Phi = \{\phi_{0}, \phi_{1}, \phi_{2}, \cdots \phi_{M}\}$, where $\phi_{0}$ denotes the trivial self metapath, $\boldsymbol{\psi}_{j j^{\prime}}$ is defined as follows: 
\begin{equation}
\label{eq:prox}
\psi_{i, j j^{\prime}}= \begin{cases}1, & \text { if } \phi_{i} \text{ exists between node $j$, $j^{\prime}$,} \\ 0, & \text { else.} \end{cases}
\end{equation}
With the attention of generating completions to the member-job ego-graph $G_{k}$ in the prompt dynamically adjusted according to Eqs. (\ref{eq:attn_adj}), (\ref{eq:prox}), the attention of PLM4Job to member/job nodes can be better aligned with the heterogeneous structure of the job marketplace.

\subsection{Task-Specific Finetuning}

The feature and structural modeling aims to encode and aggregate member/job textual features and proximity information in the job marketplace into the member/job token embeddings, such that PLM4Job can understand the member-job heterogeneous ego-graphs. In this part, we introduce the task-specific finetuning strategies for PLM4Job to generalize it for various downstream tasks.

\subsubsection{\textbf{Node-Level Tasks}}

When conducting node-level tasks on the job marketplace graph $G$ (e.g., member skill/work mode preference prediction), we first form an ego-graph-based prompt $\mathbf{x}^{e}_{k}$ with the target node $k$ as the center nodes, which includes the instruction part $\mathbf{x}^{ins}$, the ego-graph part $\mathbf{x}^{g}_{k}$, the textual features of the center node part $\mathbf{x}^{e, f}_{k}$, and the question part $\mathbf{x}^{e, q}_{k}$ as follows:
\vspace{1mm}
\begin{mdframed}[backgroundcolor=black!10] 
\textbf{Node-Level Task - Ego-Graph-based Prompt $\mathbf{x}^{e}_{k}$:} \\
\vspace{-2.5mm}

\noindent \colorbox{green!50}{\texttt{\footnotesize Given an ego-network in a job marketplace: }} \footnotesize($\mathbf{x}^{ins}$) \ \texttt{\footnotesize {\colorbox{red!35}{<member\_$k$>}} \\ 
{\colorbox{blue!35}{<member\_$i$> <member\_$j$>}} {\colorbox{yellow!50}{<job\_$l$> <job\_$m$>}}
{\colorbox{blue!35}{<member\_$o$>}} \footnotesize($\mathbf{x}^{g}_{k}$),\\ \colorbox{green!50}{the biography of the center member} \colorbox{red!35}{<member\_$k$>} \colorbox{green!50}{is:} \\
 \colorbox{green!50}{A hard-working applied research scientist at LinkedIn}} ($\color{red}\mathbf{x}^{e, f}_{k}$)\\
 \colorbox{green!50}{\texttt{The member could possess the following skills:}}
 $(\mathbf{x}^{e, q}_{k})$
\end{mdframed}
\vspace{2mm}
Here, since node-level prediction focuses more on the node feature itself, we include textual feature into the prompt $\mathbf{x}^{e}_{k}$, i.e., $\color{red}\mathbf{x}^{e, f}_{k}$. We only use member biography as an example, where other features such as member educational experience can be easily included. 

We first embed the prompt $\mathbf{x}^{e}_{k}$ with the PLM and obtain the last-layer last-step hidden representation $\mathbf{h}^{e, (-1)}_{k, -1}$. Since directly generating the target class in natural language based on $\mathbf{h}^{e, (-1)}_{k, -1}$ via autoregression may lead to hallucination \cite{huang2023survey}, e.g., outputting skills that are not in LinkedIn's standardized skill set, we introduce class tokens with embeddings $\mathbf{C}^{n} \in \mathbb{R}^{N_{C} \times d}$, where $N_{C}$ is the number of classes, and predict the label of center node $k$ as follows:
\begin{equation}
\label{eq:node_cls}
\hat{y}_{k} \sim \mathrm{Categorical}\left(softmax\left(\mathbf{C}^{n} \cdot \mathbf{h}^{e, (-1)}_{k, -1}\right)\right).
\end{equation}
For binary classification tasks, we could change the question part of the ego-graph-based prompt $\mathbf{x}^{e}_{k}$, i.e., $\mathbf{x}^{e, q}_{k}$, to \texttt{``does the member possess the skill \{name\}''}. In this case, we have two class embeddings in $\mathbf{C}^{n}$ denoting the positive and negative predictions. We directly optimize the node class embeddings $\mathbf{C}^{n}$ via Eq. (\ref{eq:node_cls}) by maximizing the log probability of the true class.

\subsubsection{\textbf{Link-Level Tasks}}

In this part, we focus on predicting one-hop relationships in the job marketplace $G$, i.e., predicting member-member following relations for \textit{people you may know (PYMK)} recommendations, and predicting member-job interactions for \textit{job you may be interested in (JYMBII)} recommendations, which form two most important business at LinkedIn. To predict the relationships, we first construct a similar ego-graph-based prompt $\mathbf{x}^{l}_{k}$ as follows:
\vspace{1mm}
\begin{mdframed}[backgroundcolor=black!10] 
\textbf{Link-Level Task - Ego-Graph-based Prompt $\mathbf{x}^{l}_{k}$:} \\
\vspace{-2.5mm}

\noindent \colorbox{green!50}{\texttt{\footnotesize Given an ego-network in a job marketplace: }} \footnotesize($\mathbf{x}^{ins}$) \ \texttt{\footnotesize {\colorbox{red!35}{<member\_$k$>}} \\ 
{\colorbox{blue!35}{<member\_$i$> <member\_$j$>}} {\colorbox{yellow!50}{<job\_$l$> <job\_$m$>}}
{\colorbox{blue!35}{<member\_$o$>}} \footnotesize($\mathbf{x}^{g}_{k}$),\\ \colorbox{green!50}{The center} \colorbox{red!35}{<member\_$k$>} \colorbox{green!50}{currently follows:} \\
 \colorbox{blue!35}{<member\_$p$> <member\_$q$> <member\_$r$>}} ($\color{red}\mathbf{x}^{l, s}_{k}$)\\
 \colorbox{green!50}{\texttt{The member may be interested following in these members:}}
 $(\mathbf{x}^{l, q}_{k})$
\end{mdframed}
\vspace{2mm}
From the above example, we can find that the difference between $\mathbf{x}^{l}_{k}$ and $\mathbf{x}^{e}_{k}$ is that, the observed end entities from the target relationship for the center node $k$, i.e., $\color{red}\mathbf{x}^{l, s}_{k}$, is included in the prompt $\mathbf{x}^{l}_{k}$. During training, we randomly mask some observed entities to form $\color{red}\mathbf{x}^{l, s}_{k}$ and stack all the hold-out neighbors as a multi-hot vector as the target $\mathbf{y}^{l}_{k} \in \{0,1\}^{\{U, I\}}$, which is generated as follows:
\begin{equation}
\label{eq:link_pred}
\hat{\mathbf{y}}^{l}_{k} \sim \mathrm{Multinomial}\left(softmax\left(\mathbf{C}^{\{U,I\}} \cdot \mathbf{h}^{l, (-1)}_{k, -1}\right)\right).
\end{equation}
Here, the weights $\mathbf{C}^{\{U,I\}}$ are the same as the weights of the LM prediction head from the first-order structural modeling for metapaths $\phi \in \{UU, UI\}$ in Eq. (\ref{eq:structure_lm}) (details see sub-section \ref{sec:struct}).

\subsection{Pipeline Summary} 

In summary, when training PLM4Job, we first pre-heat the model by optimizing Eqs. (\ref{eq:feature_lm}), (\ref{eq:structure_lm}) for $N_{pre}$ epochs. We then introduce the task-specific finetuning objective and train PLM4Job in an interleaving manner with Eq. (\ref{eq:feature_lm}), Eq. (\ref{eq:structure_lm}), and Eq. (\ref{eq:node_cls})/(\ref{eq:link_pred}). Through this strategy, both member/job textual features and heterogeneous graph structure can be fully utilized to model the job marketplace.

\subsection{Prediction}

In the prediction phase of PLM4Job, we first randomly sample $N_{g}$ ego-graphs to construct the prompt $\mathbf{x}^{\{e,l\}}_{k'}$ for the target node $k'$. We then calculate the categorical/multinomial probability according to Eq. (\ref{eq:node_cls})/(\ref{eq:link_pred}) for node/link-level tasks and take the average. Finally, for node-level tasks, we use the class token with max probability as the prediction, whereas for link prediction tasks, we rank the multinomial likelihood and suggest the top $M$ as the candidates.

\section{Experiments}

\begin{table}[t]
  \caption{Statistics of the LinkedIn job marketplace.}
  \label{tab:datasets}
 \vspace{-3mm}
  \begin{tabular}{lcccc}
    \toprule
    Dataset & \#Mem & \#Job & \#Mem-Job & \#Mem-Mem\\
    \midrule
    LinkedIn     & 69,716 & 63,368  & $490,768$  & 827,844 \\
  \bottomrule
\end{tabular}
\end{table}

\subsection{Dataset Establishment}

The studied job marketplace heterogeneous graph is established by sampling from one-day interactions between members and jobs from the United States at LinkedIn, where members' clicks, views, and applications of the job are recorded as the member-job edge in the graph under the relation "be interested in." In addition, members are connected if they work at the same company, representing the relation of "co-working." Textual attributes of the members include the headline (i.e., a brief intro. of the member under the name and photo of the member) and the biography. Textual features of the jobs include the title of the job, the company that posts the job, the job descriptions, and the skills required by the job. We collect the members' skills and work mode preferences to evaluate the node-level prediction ability of PLM4Job. Furthermore, for link-level tasks, we test the ability of PLM4Job to predict both member-job and member-member relations. The statistics of the established job marketplace heterogeneous graph are summarized in Table \ref{tab:datasets}.

\subsubsection{\textbf{Implementation Details}}
Since the decisions on the job marketplace need to be fast at LinkedIn, we use a comparatively small PLM, i.e., GPT-2 \cite{radford2018improving} with 768-dimensional token embeddings and vocabulary size of 50,257, as the PLM backbone for PLM4Job. For the metapath-based structural modeling (see Section \ref{sec:struct}), we select six metapaths $\Phi = \{UI, UU, IU, UIU, UUI, IUI\}$, where in each epoch, we randomly select one of the one-hop meta-paths and one of the two-hop meta-paths for the structural information aggregation. During the training stage, we first optimize the newly introduced ego-graph token embeddings (see Eq. (\ref{eq:token_embds})) via self-supervised feature/structural modeling as with Eqs. (\ref{eq:feature_lm}) and (\ref{eq:structure_lm}) for ten epochs to warm up the model. Then, we add the task-specific finetuning objective to subsequent epochs, where we alternately train the PLM4Job model according to Eq. (\ref{eq:feature_lm}), Eq. (\ref{eq:structure_lm}),  Eq. (\ref{eq:node_cls})/(\ref{eq:link_pred}) for 100 epochs. For the node-level tasks, we randomly select 15\% nodes with labels as the validation set and another 15\% for testing, where accuracy and F1-score are used as the metrics. For the link-level tasks, we evaluate the PLM4Job on nodes with more than five target links, where for each of such nodes, 60\% of the links are included for training, 20\% are held out for validation, and another 20\% for testing, where ranking-based metrics such as Recall@$M$ and NDCG@$M$ are used to measure the performance.

\subsection{Baselines}
\label{sec:baselines}

We compare PLM4Job with various baselines on different downstream tasks on the job marketplace. Specifically, the baselines used in this paper can be categorized into three classes: \textbf{\textit{(i)}} graph neural network (GNN)-based methods, such as GCN \cite{kipf2016semi}, GAT \cite{velivckovic2017graph}, as well as GNNs specifically designed for heterogeneous graphs, such as the heterogeneous GNN (HetGNN) \cite{hu2020heterogeneous} and heterogeneous graph attention network (HAN) \cite{wang2019heterogeneous}; \textbf{\textit{(ii)}} graph transformer-based methods such as the graphormer (GT) \cite{ying2021transformers}, the ego-graph-based transformer, Gophormer \cite{zhao2021gophormer} and the heterogeneous graph transformer (HGT) \cite{hu2020heterogeneous}; \textbf{\textit{(iii)}} the PLM-based method, i.e., InstructGCL \cite{ye2023natural}. In addition, we introduce two more baselines, i.e., SGL-Text \cite{wu2021self} and JMMFR (graph-based) \cite{zhu2022remote} for node-level tasks, and LightGCN (graph-based) \cite{he2020lightgcn} and P5 (PLM-based) \cite{geng2022recommendation} for link-level tasks.

\subsection{Node-Level Tasks}

In this subsection, we show the experiments of PLM4Job on node-level tasks on the LinkedIn job marketplace. Specifically, we are interested in two tasks, i.e., \textbf{member skill prediction}, which aims to predict whether a member has coding-related or management-related skills, and \textbf{work mode preference prediction}, which aims to predict whether a member is willing to take an online/onsite job. Since a member can have multiple skills and prefer multiple types of work modes, we model them as different binary classification problems. Both of these can significantly benefit the member-job matching at LinkedIn for better job recommendation results.
\begin{table}[t]
\setlength{\tabcolsep}{2pt}
\centering
\caption{Comparison between PLM4Job and baselines on the node-level tasks on LinkedIn job marketplace modeling.  }
\vspace{-2mm}
\label{tab:node_result}
\small
\begin{tabular}{lcc|cc}
\toprule
\textbf{Skill}&\multicolumn{2}{c|}{\textbf{Coding-Related}}&\multicolumn{2}{c}{\textbf{Manage-Related}} \\
\textbf{Dataset} & Accuracy & F1-score & Accuracy & F1-score \\ 
\midrule
MLP   & 0.7578 & 0.6919 & 0.6271 & 0.5405 \\ 
GCN \cite{kipf2016semi} & 0.7810 & 0.7263 & 0.6784 & 0.6026\\
GAT \cite{velivckovic2017graph} & 0.8053 & 0.7318  & 0.6828 & 0.5950\\
SGL-Text \cite{wu2021self} & 0.7900 & 0.7254 & 0.6933 & 0.6148\\
JMMFR \cite{zhu2022remote} & 0.8046 & \underline{0.7469} & \underline{0.7008} & 0.6196 \\
HetGNN \cite{zhang2019heterogeneous}  & \underline{0.8129} & 0.7435  & 0.6952 & 0.6169\\  
HAN \cite{wang2019heterogeneous}  & 0.8084 & 0.7280 & 0.6990 & \underline{0.6284}   \\
\midrule
GT \cite{ying2021transformers} & 0.7733 & 0.7264 & 0.6902 & 0.5997 \\
HGT \cite{hu2020heterogeneous}        & 0.7859 & 0.7315 & 0.6914 & 0.6036 \\
Gophormer \cite{zhao2021gophormer}    & 0.7925 & 0.7328 & 0.6805 & 0.6273 \\
InstructGCL \cite{ye2023natural}  & \underline{0.8103} & \underline{0.7490} & \underline{0.7052} & \underline{0.6344} \\
\midrule
PLM4Job   & \textbf{0.8187} & \textbf{0.7568} & \textbf{0.7187} & \textbf{0.6459} \\
\bottomrule
\\
\toprule
\textbf{Pref.}&\multicolumn{2}{c|}{\textbf{Onsite Jobs}}&\multicolumn{2}{c}{\textbf{Online Jobs}} \\
\textbf{Dataset} & Accuracy & F1-score & Accuracy & F1-score \\ 
\midrule
MLP   & 0.5559 & 0.4368 & 0.5746 & 0.4871 \\ 
GCN \cite{kipf2016semi} & 0.6048 & 0.5104 & \underline{0.6802} & 0.5535 \\
GAT \cite{velivckovic2017graph} & 0.5713 & 0.5091 & 0.6679 & 0.5460 \\
SGL-Text \cite{wu2021self} & 0.5825 & 0.5078 & 0.6750 & 0.5589 \\
JMMFR \cite{zhu2022remote} & 0.6094 & 0.5236 & 0.6696 & 0.5791\\
HetGNN \cite{zhang2019heterogeneous}  & \underline{0.6100} & \underline{0.5287} & 0.6737 & 0.5641 \\ 
HAN \cite{wang2019heterogeneous}  &  0.6084 & 0.5112 & 0.6784 & \underline{0.5826}\\
\midrule
GT \cite{ying2021transformers} & 0.6035 & 0.5120 & 0.6203 & 0.5477\\
HGT \cite{hu2020heterogeneous} & 0.6051 & 0.5139 & 0.6417 & 0.5582 \\
Gophormer \cite{zhao2021gophormer}  & 0.5996 & 0.5010 & 0.6629 & 0.5714 \\
InstructGCL \cite{ye2023natural}      & \underline{0.6187} & \underline{0.5245}  & \underline{0.6821} & \underline{0.5859} \\
\midrule
PLM4Job   & \textbf{0.6312} & \textbf{0.5393} & \textbf{0.6937} & \textbf{0.5924} \\
\bottomrule
\end{tabular}
\vspace{-5.5mm}
\end{table}

\subsubsection{\textbf{Comparison with Baselines}}

We first compare the proposed PLM4Job with the baselines introduced in Section \ref{sec:baselines}, where the results are summarized in Table \ref{tab:node_result}. From Table \ref{tab:node_result}, we can find that heterogeneous GNNs generally show better performance than the normal GNN models due to their explicit consideration of different relations in the job marketplace graph. However, since these models use bag-of-word representations to model member/job textual features, their shallow understanding of important textual features leads to overall unsatisfactory results. For the graph transformer (GT)-based methods, HGT can distinguish heterogeneous relationships in the job market graph, but as a global model, it may not be able to fully utilize the local information for predictions. Gophormer is specifically designed for ego-graphs, but it does not consider the heterogeneous structure in the job marketplace. Most importantly, although GT-based methods have a similar underlying transformer structure as the PLM, these models are not pre-trained on large datasets and do not contain prior knowledge of the natural language. Therefore, their understanding of the member/job textual features as well as their relationship in the job marketplace is also shallow. As a PLM-based graph mining algorithm, InstructGCL performs the best among all the baselines as it utilizes the pretrained knowledge of PLMs. However, it does not consider the heterogeneous relationships in the job market graph. In addition, the proximity information is described via natural language such as "one-hop," etc., which may not faithfully direct the attention of the PLM according to the proximity information in the heterogeneous job marketplace ego-graph. In contrast, by tightly coupling the heterogeneous local structure of job marketplace graph with the pre-trained knowledge of the PLM, the proposed PLM4Job achieves the best results on the four datasets across all the metrics.

\begin{table}[t]
\setlength{\tabcolsep}{2pt}
\centering
\caption{Ablation study for PLM4Job on node-level tasks.}
\label{tab:node_ab}
\small
\begin{tabular}{lcc|cc}
\toprule
\textbf{Skill}&\multicolumn{2}{c|}{\textbf{Coding-Related}}&\multicolumn{2}{c}{\textbf{Manage-Related}} \\
\textbf{Dataset} & Accuracy & F1-score & Accuracy & F1-score \\ 
\midrule
PLM4Job-NE  & 0.8129 & 0.7501 & 0.7090 & 0.6422\\
PLM4Job-NA  & 0.8094 & 0.7335 & 0.7046 & 0.6368 \\
PLM4Job-N2  & 0.8073 & 0.7274 & 0.6813 & 0.6237\\
\midrule
PLM4Job   & \textbf{0.8187} & \textbf{0.7568} & \textbf{0.7187} & \textbf{0.6459} \\
\bottomrule
\\
\toprule
\textbf{Pref.}&\multicolumn{2}{c|}{\textbf{Onsite Jobs }}&\multicolumn{2}{c}{\textbf{Online Jobs}} \\
\textbf{Dataset} & Accuracy & F1-score & Accuracy & F1-score \\ 
\midrule
PLM4Job-NE  & 0.6281 & 0.5336 & 0.6841 & 0.5809\\
PLM4Job-NA  & 0.6248 & 0.5312 & 0.6786 & 0.5781 \\
PLM4Job-N2  & 0.6125 & 0.5237 & 0.6760 & 0.5754 \\
\midrule
PLM4Job   & \textbf{0.6312} & \textbf{0.5393} & \textbf{0.6937} & \textbf{0.5924} \\
\bottomrule
\end{tabular}
\vspace{-2mm}
\end{table}

\subsubsection{\textbf{Ablation Studies}}
\label{sec:ab_nodes}

In this part,  we conduct ablation study to show the effectiveness of the ego-graph-based prompt (see Section \ref{sec:prompt}) and the proximity-aware attention alignment strategy (see Section \ref{sec:proxi}). Specifically, three ablation models are introduced on PLM4Job, where \textit{PLM4Job-NE} removes the entity and graph positional embeddings, \textit{PLM4Job-NA} removes the proximity-aware attention alignment module, \textit{PLM4Job-N2} removes the second-order meta-paths in structural modeling. The results are summarized in Table \ref{tab:node_ab}. For Table \ref{tab:node_ab}, we can find that proximity-based attention alignment contributes significantly to the superior performance of PLM4Job, which demonstrates the misalignment of the attention original PLM with the proximity relations in the heterogeneous graph structure. In addition, the combination of entity and ego-graph positional embeddings facilitates PLM4Job to well distinguish different nodes in the heterogeneous job marketplace ego-graph. 

\subsection{Link-Level Tasks}
In this sub-section, we show the experimental results of link-level prediction tasks on the LinkedIn job marketplace. Specifically, we focus on predicting member-job interactions (i.e., JYMBII prediction) and member-member interactions (i.e., PYMK prediction).

\begin{table}[t]
\setlength{\tabcolsep}{2pt}
\centering
\caption{Comparison between PLM4Job and various baselines on the link-level tasks for job marketplace modeling. }
\label{tab:link_result}
\small
\begin{tabular}{lccc}
\toprule
\textbf{Link}&\multicolumn{3}{c}{\textbf{Member-Job Interaction}} \\
\textbf{Dataset} & Recall@20 & Recall@40 & NDCG@100  \\ 
\midrule
Dual-MLP   & 0.0819 & 0.1318 & 0.0703 \\ 
GCN \cite{kipf2016semi} & 0.1280 & 0.1905 & 0.0946\\
GAT \cite{velivckovic2017graph} & 0.1204 & 0.1828 &  0.0912 \\ 
LightGCN \cite{he2020lightgcn} &0.1351 & \underline{0.1985} & 0.1027\\
HetGNN \cite{zhang2019heterogeneous}  & 0.1331 & 0.1957 & 0.1025\\ 
HAN \cite{wang2019heterogeneous}  & \underline{0.1386} & 0.1971 & \underline{0.1089}\\
\midrule
GT \cite{ying2021transformers} & 0.1013 & 0.1830 & 0.0976\\
HGT \cite{hu2020heterogeneous} & 0.1206 & 0.1929 & 0.0970\\
Gophormer \cite{zhao2021gophormer}  & 0.1173 & 0.1852 &  0.0947\\
P5 \cite{geng2022recommendation} & 0.1412 & 0.2116 & 0.1049\\
InstructGCL \cite{ye2023natural}  & \underline{0.1437} & \underline{0.2158}  & \underline{0.1104} \\
\midrule
PLM4Job   & \textbf{0.1545} & \textbf{0.2210} & \textbf{0.1153} \\
\bottomrule
\\
\toprule
\textbf{Link}&\multicolumn{3}{c}{\textbf{Member-Member Interaction}} \\
\textbf{Dataset} & Recall@20 & Recall@40 & NDCG@100  \\ 
\midrule
Dual-MLP   & 0.1199 & 0.1843 & 0.1085 \\ 
GCN \cite{kipf2016semi} & 0.1484 & 0.2326 & 0.1421 \\
GAT \cite{velivckovic2017graph} & 0.1577 & 0.2490 &  0.1578\\
LightGCN \cite{he2020lightgcn} & \underline{0.1584} & \underline{0.2501} & \underline{0.1597} \\
HetGNN \cite{zhang2019heterogeneous}  & 0.1462 & 0.2285 & 0.1402 \\ 
HAN \cite{wang2019heterogeneous}  & 0.1513 & 0.2371 & 0.1456 \\
\midrule
GT \cite{ying2021transformers} & 0.1437 & 0.2312 & 0.1483\\
HGT \cite{hu2020heterogeneous} & 0.1470 & 0.2433 & 0.1537\\
Gophormer \cite{zhao2021gophormer}  & 0.1526 & 0.2498 & 0.1584\\
P5 \cite{geng2022recommendation} & \underline{0.1801} & \underline{0.2575} & 0.1695\\
InstructGCL \cite{ye2023natural}    & 0.1795 & 0.2531 & \underline{0.1762} \\
\midrule
PLM4Job   & \textbf{0.1953} & \textbf{0.2684} & \textbf{0.1809} \\
\bottomrule
\end{tabular}
\vspace{-3mm}
\end{table}

\subsubsection{\textbf{Comparison Results}}

Similarly, we first compare PLM4Job with various GNN-based, GT-based, and PLM-based baselines, where the results are summarized in Table \ref{tab:link_result}. From Table \ref{tab:link_result}, we can find that, generally, PLM4Job outperforms most of the GNN/GT/PLM-based baselines, which demonstrates its ability to generalize to link prediction tasks on the job marketplace. In addition, ablation studies are also conducted for the link prediction task, where the introduced ablation models are the same as the ones used in subsection \ref{sec:ab_nodes}. The results are summarized in Table \ref{tab:link_ab}. From the Table, we can find that all components of the proposed PLM4Job also contribute positively to its final superior results for link-level tasks.

\subsection{\textbf{Deployment of PLM4Job Embeddings}}

PLM4Job intends to serve as the foundation model for the LinkedIn job retrieval system. At LinkedIn, the L1 retrieval model is evaluated from the user feedback on the L2 ranking model. Since PLM4Job is expensive to deploy directly, we extract the member/job token embeddings from the trained PLM4Job model (takes O(1) complexity), reduce their dimension, and deploy them on two of the most important systems at LinkedIn: \textit{JYMBII} and \textit{PYMK}. We name the two-tower model with PLM4Job embeddings as PLM4Job-Emb.

We compare the PLM4Job-Emb model with another model that adds the embeddings of M6-Rec \cite{cui2022m6}, i.e., a PLM-based matching method for recommendations (where we denote the model as M6-Rec-Emb), as well as the original two tower model. Specifically, we randomly split the members into three folds and evaluate the three models on users' feedback on the L2 ranking model accumulated in a week. From Table \ref{tab_pymk}, we can find that, adding PLM4Job member/job embeddings can improve the performance of the existing two-tower model at LinkedIn (which includes embeddings from internally trained BERT \cite{kenton2019bert} and GNNs), which further demonstrates the ability of PLM to serve as foundation models for job marketplace and adapt to downstream tasks with effectiveness and efficiency.

\section{Conclusion}

In this paper, we proposed PLM4Job, a graph-oriented pre-trained language model, to serve as the foundation model for job marketplace modeling. Specifically, we first propose an ego-graph-based prompt to facilitate the PLM to understand the features, relations, and local structure of the job marketplace with the pretrained knowledge. In addition, a proximity-aware attention alignment strategy is proposed to align the attention of the PLM with the heterogeneous proximity relations among members and jobs in the job marketplace ego-graph. Extensive experiments on LinkedIn real-world data demonstrate the effectiveness of PLM4Job.

\section*{Acknowledgment}

This work is supported in part by the National Science Foundation under grants (IIS-2006844, IIS-2144209, IIS-2223769, CNS-2154962, BCS-2228534, and CMMI-2411248), and the Commonwealth Cyber Initiative Awards under grant (VV-1Q24-011).
\begin{table}[t]
\setlength{\tabcolsep}{2pt}
\centering
\caption{Ablation Study for PLM4Job on link-level tasks. }
\label{tab:link_ab}
\small
\begin{tabular}{lccc}
\toprule
\textbf{Link}&\multicolumn{3}{c}{\textbf{Member-Job Interaction}} \\
\textbf{Dataset} & Recall@20 & Recall@40 & NDCG@100  \\ 
\midrule
PLM4Job-NE  & 0.1502 & 0.2188 & 0.1131 \\
PLM4Job-NA  & 0.1477 & 0.2095 & 0.1046 \\
PLM4Job-N2  & 0.1410 & 0.2163 & 0.1094 \\
\midrule
PLM4Job   & \textbf{0.1545} & \textbf{0.2210} & \textbf{0.1153} \\
\bottomrule
\\
\toprule
\textbf{Link}&\multicolumn{3}{c}{\textbf{Member-Member Interaction}} \\
\textbf{Dataset} & Recall@20 & Recall@40 & NDCG@100  \\ 
\midrule
PLM4Job-NE  & 0.1927 & 0.2675 & 0.1798 \\
PLM4Job-NA  & 0.1761 & 0.2503 & 0.1716 \\
PLM4Job-N2  & 0.1809 & 0.2517 & 0.1753 \\
\midrule
PLM4Job   & \textbf{0.1953} & \textbf{0.2684} & \textbf{0.1809} \\
\bottomrule
\end{tabular}
\end{table}

\begin{table}[t]
\setlength{\tabcolsep}{2pt}
\centering
\caption{Deploy PLM4Job embeddings to the two-tower models on the LinkedIn JYMBII and PYMK systems.}
\small
\begin{tabular}{lccc}
\toprule
\textbf{Online}&\multicolumn{3}{c}{\textbf{Member-Job Interaction (JYMBII)}} \\
\textbf{Dataset} & Recall@20 & Recall@40 & NDCG@100  \\ 
\midrule
Two-Tower  &  0.1353 & 0.1937 & 0.0980 \\
M6-Rec-Emb  & 0.1409 & 0.2062 & 0.1058 \\
\midrule
PLM4Job-Emb  & \textbf{0.1426} & \textbf{0.2099} & \textbf{0.1093} \\
\bottomrule
\\
\toprule
\textbf{Online}&\multicolumn{3}{c}{\textbf{Member-Member Interaction (PYMK)}} \\
\textbf{Dataset} & Recall@20 & Recall@40 & NDCG@100  \\ 
\midrule
Two-Tower  &  0.1475 & 0.2360 & 0.1437 \\
M6-Rec-Emb  & 0.1505 & 0.2374 & 0.1481 \\
\midrule
PLM4Job-Emb  & \textbf{0.1593} & \textbf{0.2446} & \textbf{0.1528}  \\
\bottomrule
\end{tabular}
\label{tab_pymk}
\vspace{-5mm}
\end{table}

\balance
\bibliographystyle{ACM-Reference-Format}
\bibliography{PLM4Job}

\end{document}